\documentclass[journal=cmatex,manuscript=article,
                layout=onecolumn]{achemso}
\usepackage[T1]{fontenc} % Use modern font encodings
\usepackage{mathtools}
\usepackage{multirow}
\usepackage{amsmath}
\usepackage{amssymb}
\usepackage{caption}
\usepackage{enumitem}
\usepackage{xcolor}
\usepackage[colorlinks=true,allcolors=blue]{hyperref}
\usepackage{bm}
\usepackage{courier}
\usepackage{times}
\usepackage{setspace}
\usepackage{threeparttable}
\newcommand{\DRtuning}{S6}
\newcommand{\fcontinuum}{S8}
\newcommand{\fmodulardesign}{S9}
\author{R. Patrick Xian}
\affiliation{Department of Engineering, University of Cambridge, Trumpington Street, \\Cambridge CB2 1PZ, United Kingdom.}
\alsoaffiliation{Department of Chemistry and Biochemistry, University of South Carolina, 631 Sumter Street, Columbia, SC 29208, USA.}
\email{xrpatrick@gmail.com}
\author{Ryan J. Morelock}
\affiliation{Department of Chemical and Biological Engineering, University of Colorado Boulder, \\3415 Colorado Avenue, Boulder, CO 80303, USA.}
\altaffiliation{Current address: Center for Nanophase Materials Science, Oak Ridge National Laboratory, \\1 Bethel Valley Road, Oak Ridge, TN 37831, USA.}
\author{Ido Hadar}
\affiliation
{The Institute of Chemistry, Casali Center for Applied Chemistry, and the Center for Nanoscience and Nanotechnology, The Hebrew University of Jerusalem, Edmond J. Safra Campus, Givat-Ram, 9190401 Jerusalem, Israel.}
\author{\\Charles B. Musgrave}
\affiliation{Department of Chemical and Biological Engineering, University of Colorado Boulder, \\3415 Colorado Avenue, Boulder, CO 80303, USA.}
\altaffiliation{Current address: John and Marcia Price College of Engineering, University of Utah,\\ 72 S. Central Campus Drive, Salt Lake City, UT 84112, USA.}
\author{Christopher Sutton}
\affiliation{Department of Chemistry and Biochemistry, University of South Carolina, 631 Sumter Street, Columbia, SC 29208, USA.}
\alsoaffiliation{Department of Materials Science and Engineering, University of Toronto, 184 College Street, Toronto, ON M5S 3E4, Canada.}
\email{ca.sutton@utoronto.ca}

\title[]
  {From structure mining to unsupervised exploration of atomic octahedral networks}

\begin{document}
% \begin{tocentry}

% Some journals require a graphical entry for the Table of Contents.
% This should be laid out ``print ready'' so that the sizing of the

% Inside the \texttt{tocentry} environment, the font used is Helvetica
% 8\,pt, as required by \emph{Journal of the American Chemical

% The surrounding frame is 9\,cm by 3.5\,cm, which is the maximum
% permitted for  \emph{Journal of the American Chemical Society}
% graphical table of content entries. The box will not resize if the
% content is too big: instead it will overflow the edge of the box.

% This box and the associated title will always be printed on a
% separate page at the end of the document.

% \end{tocentry}

%% 4000 characters

\newpage
% \begin{abstract}
% \vspace{em}
\begin{quote}
\setstretch{1.3}
\bf{Understanding the spatial arrangements of atom-centered coordination octahedra is crucial for relating structures to properties for many materials families. Traditional case-by-case inspection becomes a prohibitive task for discovering trends and similarities in large datasets. Here, we operationalize chemical intuition to automate the geometric parsing, quantification, and classification of coordination octahedral networks using unsupervised machine learning. We apply the workflow to analyze two datasets to demonstrate its effectiveness. For computationally generated single oxide perovskite (ABO$_{3}$) polymorphs, we uncover axis-dependent tilting trends which assist in detecting oxidation state changes. For hybrid iodoplumbates (A$_x$Pb$_y$I$_z$) from measured structures, we taxonomize their octahedral networks, revealing a Pauling-like connectivity rule for the coordination environment and the design principles underpinning their structural diversity. Our results offer a glimpse into the vast design space of atomic octahedral networks in materials chemistry and inform high-throughput, targeted screening of specific structure types.}
\end{quote}
% \end{abstract}
\begin{figure*}[htbp!]
    \centering
    \includegraphics[width=0.9\linewidth]{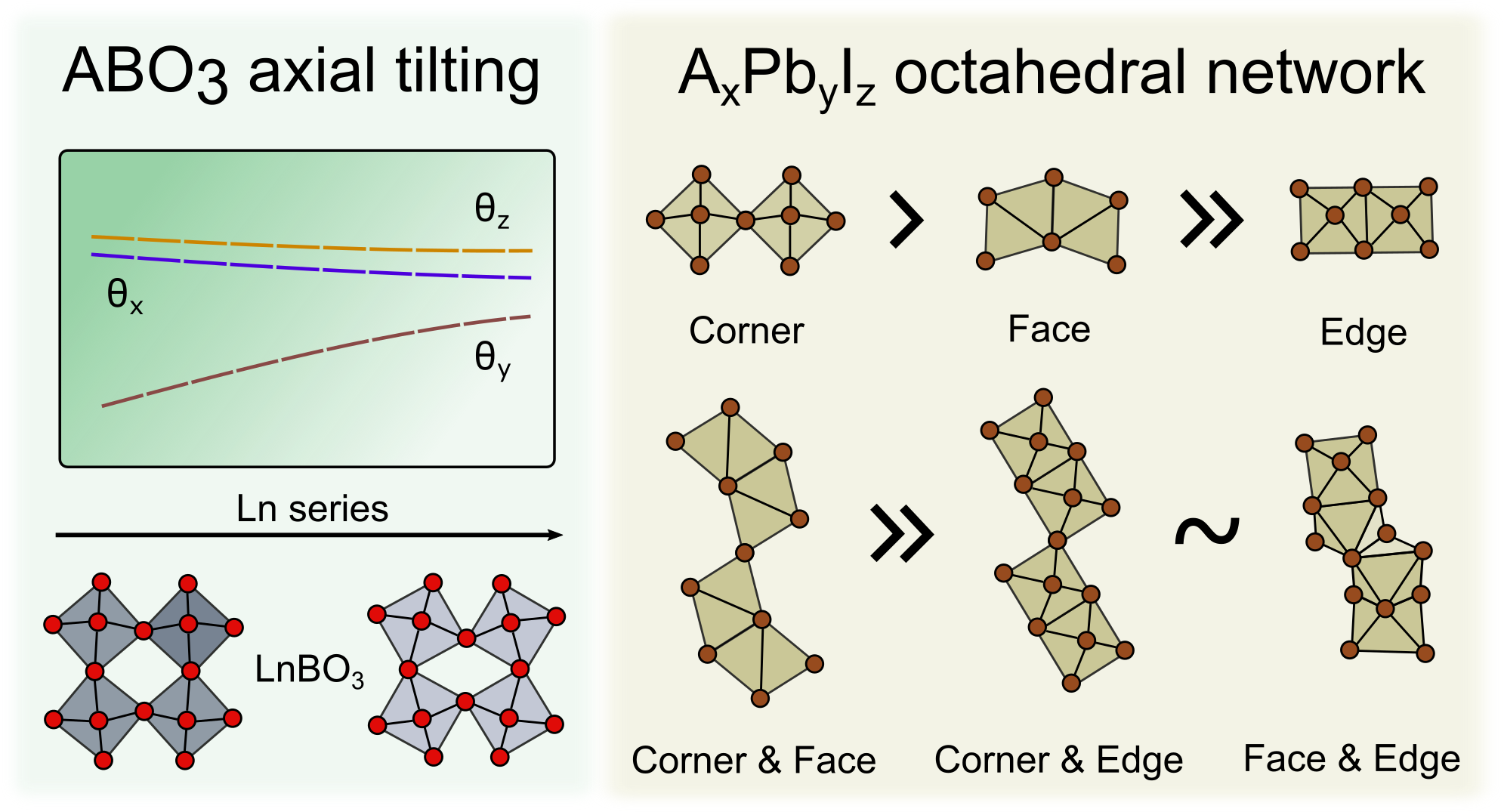}
    \begin{flushleft}
        Highlights of the two case studies in this work on the computational analysis of materials distinguished by their octahedral networks. Left: Axis-resolved tilting trends from A-site substitution in lanthanide oxide perovskite series (LnBO$_3$). Right: Pauling-like rules for the prevalence of connectivity patterns among the inorganic polytypes of hybrid iodoplumbates (A$_x$Pb$_y$I$_z$).
    \end{flushleft}
    \label{fig:toc}
\end{figure*}

\setstretch{1.23}
\section{Introduction}
The coordination polyhedron is a recurring theme in chemistry from molecules to solids \cite{Pauling1929,Wells1977,Ferey2000,Ferraris2008,Hawthorne2014,Bindi2020}. Among inorganic compounds, coordination octahedra (COs) are one of the most common local atomic environments of metal-centered ions \cite{Waroquiers2017}. Octahedral networks (ONs, i.e., networks of COs) display elaborate spatial patterns based on orientation (i.e., tilting), distortion, and connectivity, which underlie the stability and functionalities of materials. The most prominent representatives of materials containing ONs belong to the perovskite structure family \cite{Aleksandrov1997}.

Besides the corner-sharing ONs in classic perovskites, their variants are prevalent in binary and multinary compounds \cite{Pauling1929,Wells1977,Gregory2001,Muller2007}, metal-rich clusters \cite{Simon1988} (with anion-centered COs), numerous hydroxide-based crystalline minerals \cite{Hawthorne2014,Bindi2020}, and functional materials \cite{Williams2006,Saparov2016,Mao2019,Bostrom2020}, etc. Incidentally, ONs are also a convenient abstraction of the underlying atomistic identity and are enumerable algorithmically \cite{Hawthorne2014}, providing an intuitive surrogate to first-principles approaches for explaining the mechanisms leading to outstanding material properties \cite{Cammarata2014,Voskanyan2021}. Therefore, it is not an understatement that cataloging and documenting the architecture and distribution of ONs greatly facilitates our understanding of matter \cite{Wells1973,Ferey2000}.

\begin{figure*}[htb!]
  \begin{center}
    \includegraphics[width=0.93\textwidth]{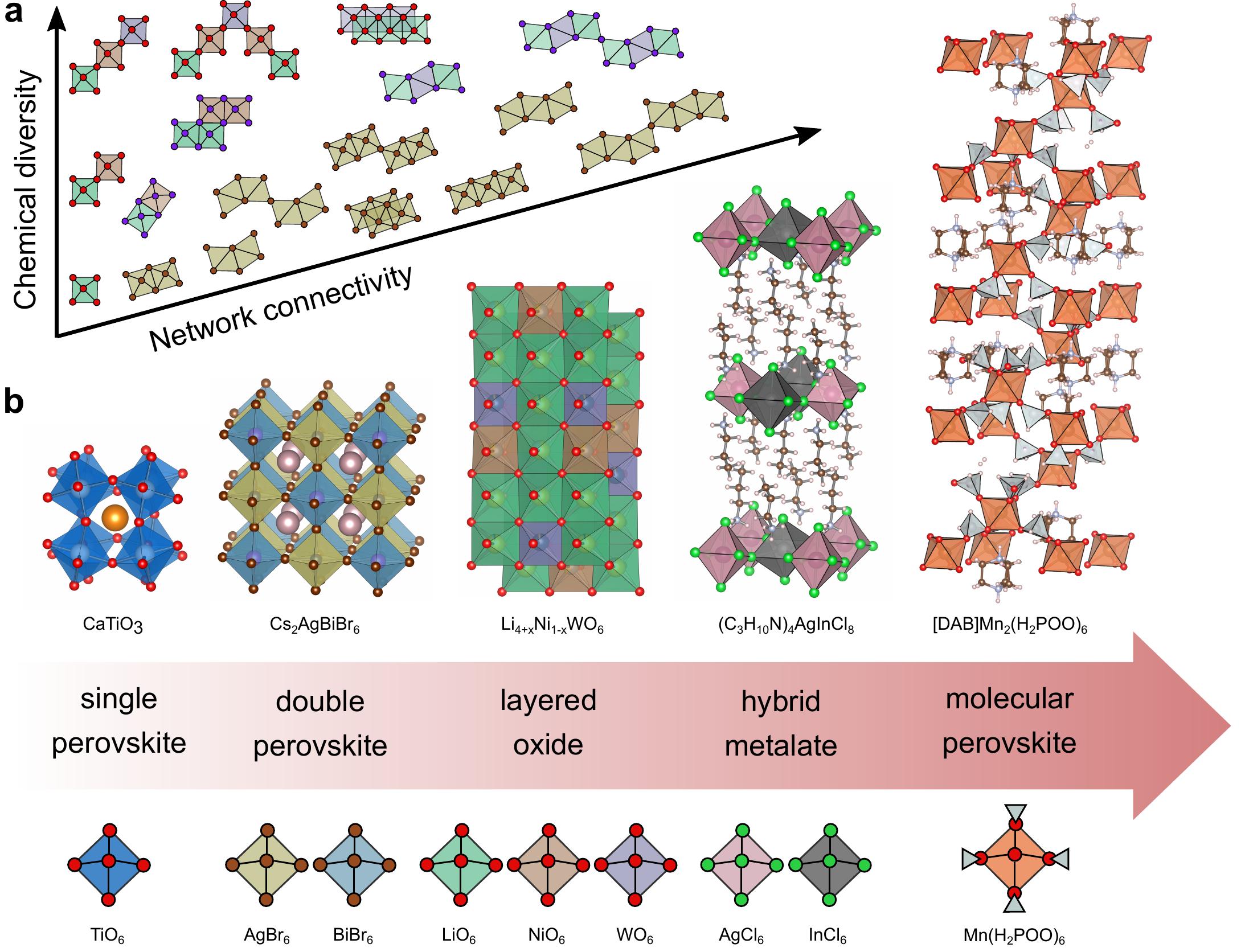}
    \caption{\textbf{Spatial complexity of octahedral networks.} \textbf{a}, Illustration of the spatial complexity of octahedral networks (ONs) along two directions: chemical diversity and network connectivity. The difference in chemical composition is represented by color. The types of connectivity between octahedra are categorized as corner-, edge-, and face-sharing. \textbf{b}, A progression of exemplary materials containing ONs is shown from left to right. The structures are taken from materials databases (single perovskite CaTiO$_3$) and publications (double perovskite Cs$_2$AgBiBr$_6$, layered nonstoichiometric oxide Li$_{4+x}$Ni$_{1 - x}$WO$_6$ \cite{Taylor2019}, hybrid multinary perovskite (C$_3$H$_{10}$N)$_4$AgInCl$_8$ \cite{Mao2019}, and molecular perovskite [DAB]Mn$_2$(H$_2$POO)$_6$ \cite{Wu2017}). The legend at the bottom indicates the types of coordination octahedra present in the corresponding crystal structures directly above each octahedron.}
    \label{fig:hierarchy}
  \end{center}
  \vspace{-1em}
\end{figure*}

Designing materials that contain ONs may be viewed through the lens of spatial patterning or tiling. Although the ON design space extends in multiple directions \cite{Hallweger2022}, in this work, we computationally examined two dimensions: chemical diversity and network connectivity, as illustrated in Fig. \ref{fig:hierarchy}a, taking the perovskite family of materials as examples. Along the chemical diversity axis, alloying multiple ions can create new properties. Single perovskites have the formula unit ABX$_3$, where A and B are cations, and X is an anion. The oxide single perovskites (ABO$_3$), which are well-studied, has a total member count approaching $5\times10^3$. This number quickly explodes to $10^7$ for double oxide perovskites (AA$^{\prime}$BB$^{\prime}$O$_6$) \cite{Filip2018,Bartel2019,bare_dataset_2023} and much more for higher-order multinary perovskites. The connectivity within the ONs may be modified by pairing with selected organic cations in these structures, which in turn alter the structural stability and electronic properties without tuning the chemical diversity of the ON. This results in the distinctive connectivity patterns that characterize the hybrid (i.e., containing both organic and inorganic parts) perovskites and their extended families of materials, called perovskitoid or hybrid metalates \cite{Saparov2016,Stoumpos2017,Smith2018,Jin2021,gilley2025}.

Existing methods for quantifying ONs in chemical structures date back at least half a century: The Glazer tilt system proposed in the 1970s \cite{Glazer1972} classifies the spatial orientation of corner-sharing COs in perovskites and is widely adopted by the community. The periodic nets introduced by Wells from the 1950s to the 1970s \cite{Wells1973,Wells1977} and the structural hierarchy hypothesis introduced by Hawthorne shortly afterwards \cite{Hawthorne2014} provide an intuitive simplification of the ONs for understanding their similarity and complexity. In recent decades, advances in chemical synthesis have greatly expanded the diversity of compounds containing ONs (see Fig. \ref{fig:hierarchy}b), which exhibit seemingly unlimited possible arrangements. Although the classification nomenclature for these modular structures is still under active debate, the materials families containing ONs are increasingly viewed as a continuum with shared chemical provenance and connectivity patterns \cite{Mercier2019,Akkerman2020}. Moreover, modern atomic-resolution electron microscopy is capable of characterizing local disordering at heterogeneous interfaces \cite{He2015}, creating opportunities to broadly sample the configurational space and cross-examine experimental results with theoretical predictions. Driven by the growing quantity of structural data and the community efforts towards developing compact and efficient representations of molecular and materials systems \cite{Onat2020,Musil2021}, we present here a computational workflow for the structural analysis and discovery of atomic ONs to obtain summary statistics and occurrence ranking. These are enabled by geometry processing \cite{Botsch2010}, unsupervised machine learning (manifold learning and clustering), and human label refinement with the use of a coordination network encoding (CNE) scheme with scale invariance. The collection of these methods reveals the octahedral tilting trends in oxide perovskite polymorphs and guides the identification of inorganic polytypes in hybrid iodoplumbates (i.e., lead iodide salts), which led us to discover the preference of connectivity in contrast with Pauling's third rule \cite{Pauling1929,George2020} in these materials and the power-law distribution of polytypes.

\section{Results}

We developed an atomic structure parsing workflow (see SI section 1) to analyze ONs using two graphs: (i) the mesh graph which uses atoms as nodes and involves the outer surface of the coordination polyhedra; (ii) the inter-unit graph which uses the COs as the nodes. The mesh graph provides connectivity information at the atomic level, while the inter-unit graph includes the connectivity information for the COs. They capture the hierarchical information in the ONs and provide the basis for analyzing large materials structure datasets involving ONs to gain insights into the design principles of complex (e.g., multivalent) perovskites and hybrid perovskite-derived materials. In the following, we present the results on oxide perovskite and hybrid iodoplumbate datasets.

\subsection{Octahedral tilting in oxide perovskite (ABO$_3$) polymorphs}

The discussions on octahedral tilting trends are normally based on the examination of specific subclasses of single perovskite structures that have been synthesized \cite{Cammarata2014,Bechtel2018}. Proxies for the tilting angle \cite{Filip2014,Ganose2019,Saidi2020}, such as the B-O-B angle, have often been used to estimate the octahedra orientation, and typically only small datasets of experimental data \cite{Straus2020} have been examined in attempts to uncover trends between these proxies and materials properties. However, a large proportion of the abundant oxide perovskites found on earth persist in kinetically trapped metastable states, either as polymorphs with free energies above the convex hull vs. composition phase diagram or as precursors or intermediates that form during chemical synthesis \cite{Sun2016}. Consequently, understanding the large design space of oxide perovskites requires evaluating many configurations, including those of metastable polymorphs.
\begin{figure*}[htb!]
  \begin{center}
    \includegraphics[width=0.96\textwidth]{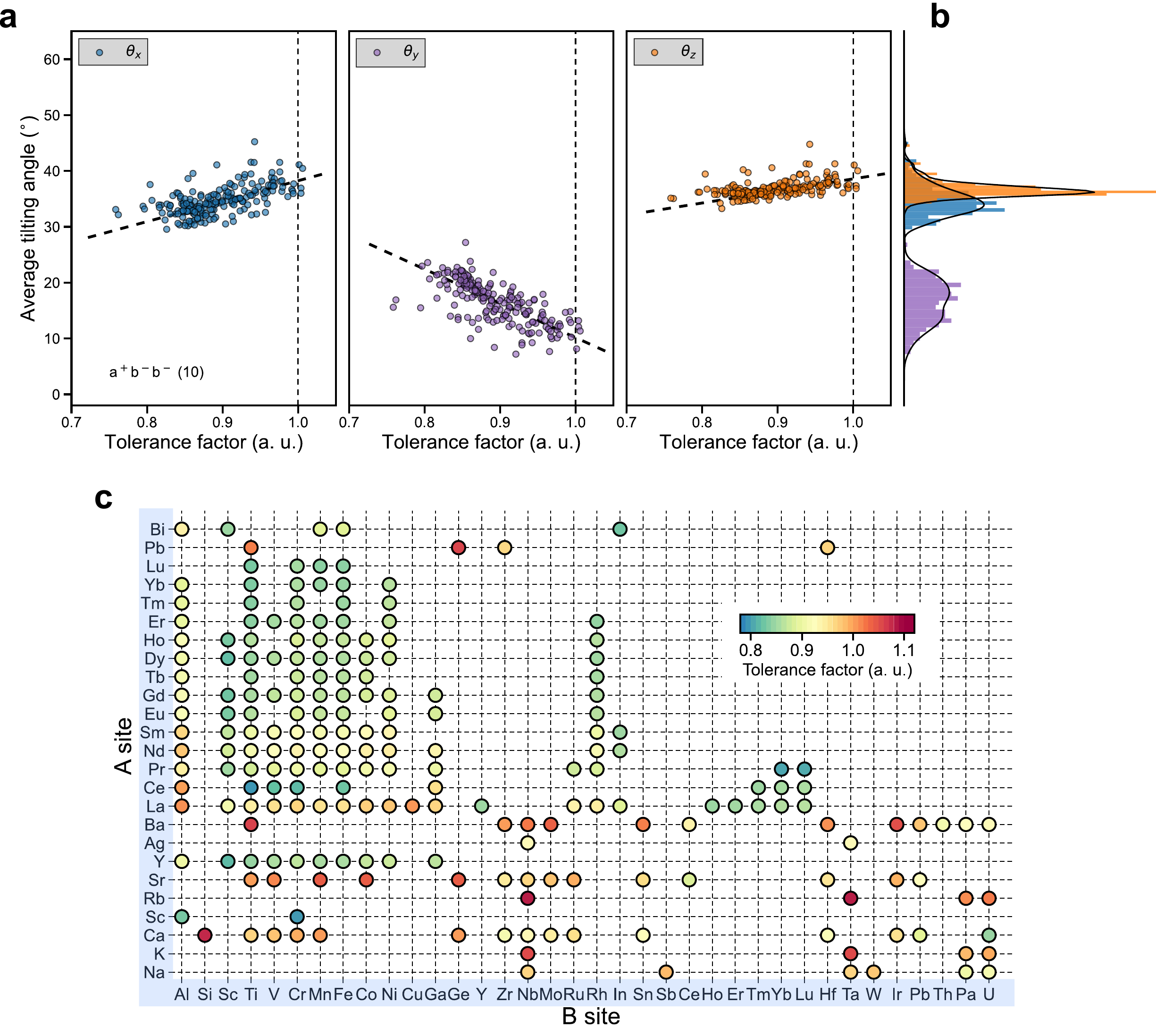}
    \caption{\textbf{Octahedral tilting trends in orthorhombic ABO$_3$ polymorphs.} \textbf{a}, Dependence on the Goldschmidt tolerance factor \cite{Goldschmidt1926} for the axis-dependent tilting angles of single oxide perovskite polymorphs within the tilt class a$^+$b$^-$b$^-$ (\#10). The dashed lines highlight the linear trend. The density estimation is shown in \textbf{b} with corresponding colors. \textbf{c}, Map of the Goldschmidt tolerance factor through the computationally accessed chemical space.}
    \label{fig:tilting_angles_1}
  \end{center}
  \vspace{-1em}
\end{figure*}

The structure mining workflow we constructed leads to a more complete picture, along with the availability of a large dataset of oxide perovskite polymorphs from high-throughput computation \cite{Bare2022} (see SI section 2). The dataset includes a total of over 2000 structures (see Methods) divided into ten different Glazer tilt classes \cite{Glazer1972}. We illustrate our results in Fig. \ref{fig:tilting_angles_1} using orthorhombic polymorphs in the a$^+$b$^-$b$^-$ (\#10) tilt class, while additional information is provided in SI section 2. Two distinctive trends stand out in Fig. \ref{fig:tilting_angles_1}: (1) There exists a clear dependence (the global trends) of tilting angles on the Goldschmidt tolerance factor \cite{Goldschmidt1926} for almost all non-cubic polymorphs; (2) When ranked by the periodicity of the elements, we observe repeating patterns (the microtrends) of the tilting angles. The observation indicates that, in certain classes of polymorphs, the atomic structures obtained through isovalent substitution of $A$-site (or $B$-site) cations are highly predictable, which gives rise to a periodic table of oxide perovskite polymorphs. These microtrends reflect the dependence of chemical substitution on the structure (or property) in a substitution series.
\begin{figure*}[htb!]
  \begin{center}
    \includegraphics[width=1\textwidth]{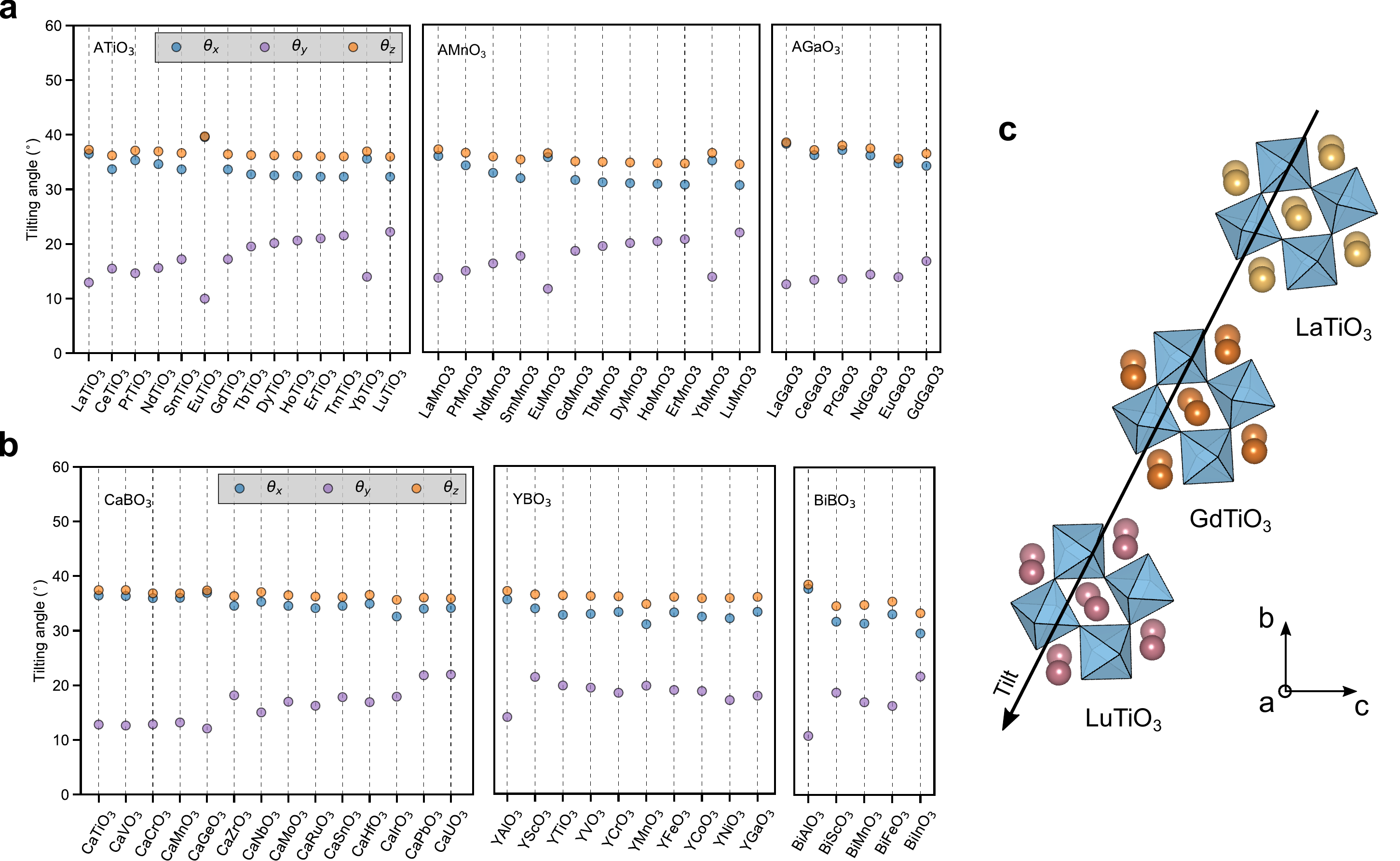}
    \caption{\textbf{Axis-dependent tilting in ABO$_3$ substitution series.} Microtrends of axis-dependent octahedral tilting found for representative $A$-site (\textbf{a}) and $B$-site (\textbf{b}) cation substitution series. \textbf{c}, An example series of lanthanide oxide perovskite structures exhibiting tilting microtrends. The three structures shown here are orthorhombic LaTiO$_3$, GdTiO$_3$, and LuTiO$_3$ with the TiO$_6$ coordination octahedra colored in light blue. The structures are displayed along the crystallographic \textbf{a} axis as indicated in the compass. }
    \label{fig:tilting_angles_2}
  \end{center}
  \vspace{-1em}
\end{figure*}

A closer look at the $A$-site substitution series (Fig. \ref{fig:tilting_angles_2}a), such as lanthanide metalates (LnBO$_3$), reveals a monotonic change in all tilting angles as the ionic radius decreases (the lanthanide contraction). These microtrends are similar in orthorhombic lanthanide titanates (LnTiO$_3$), manganites (LnMnO$_3$), and gallates (LnGaO$_3$). The structural changes also correlate with the changes in their experimental properties as previously reported \cite{Goodenough2004,Gschneidner2016}. In $B$-site substitution series (see Fig. \ref{fig:tilting_angles_2}b), a similar pattern emerges, albeit across a broader range of substituted cations. These tilting microtrends within a large dataset can identify outlier materials directly based on their structural deviations. In the case of oxide perovskites, we find that the europium and ytterbium perovskites break the tilting trends (see Fig. \ref{fig:tilting_angles_2}a). Subsequent charge analysis of all cations (see SI section 2) in the orthorhombic LnTiO$_3$ and LnMnO$_3$ series using DFT+U onsite magnetic moments show that europium and ytterbium ions are in the +2 oxidation state, forming divalent-tetravalent perovskites (A$^{2+}$B$^{4+}$O$_3$), whereas the remaining compounds in the series are trivalent perovskites (A$^{3+}$B$^{3+}$O$_3$). Typically, assigning oxidation states in large computational screening datasets is a complicated task, but our analysis shows that the structural trend obtained from COs is a convenient indicator for changes in the oxidation state of cations.

\subsection{Octahedral network motifs in hybrid iodoplumbates (A$_x$Pb$_y$I$_z$)}
Small organic molecules have been traditionally used to template the ON \cite{Mitzi2001}, resulting in a wide range of modular (or recombination) structures \cite{Lima-de-Faria1990,Ferraris2008} with a multimeric ON as the inorganic framework. A multimeric ON contains more than a single CO in its motif. Besides complex oxides \cite{Krivovichev1997}, hybrid metalates \cite{Jin2021} such as iodoplumbates are prominent examples. Existing terminologies used to describe the ON connectivity pattern are unstructured. For example, purely corner-sharing network motifs are classified by (i) the cleaving plane index \cite{Mitzi2001,Saparov2016}, (ii) homologous series \cite{Cao2015,Soe2019}, which contain stacked octahedral layers of varying thickness, and (iii) corrugated network \cite{Mao2017}, which contains W-shaped octahedral sequences with mixed \textit{cis} and \textit{trans} neighbor configurations. These classification conventions are mostly derived from the crystallographic shear structures \cite{Voskanyan2021}. Moreover, the umbrella term \textit{perovskitoid} from the halide perovskite community \cite{Stoumpos2017,gilley2025} was originally used to denote 1D structures with face-sharing octahedra, but now generally refers to the extended family of hybrid metalates without purely corner-sharing ONs. In general, these attempts lack specificity in capturing the variety of low-dimensional structures and are incompatible with high-throughput computational screening.
\begin{figure*}[htb!]
  \begin{center}
    \includegraphics[width=\textwidth]{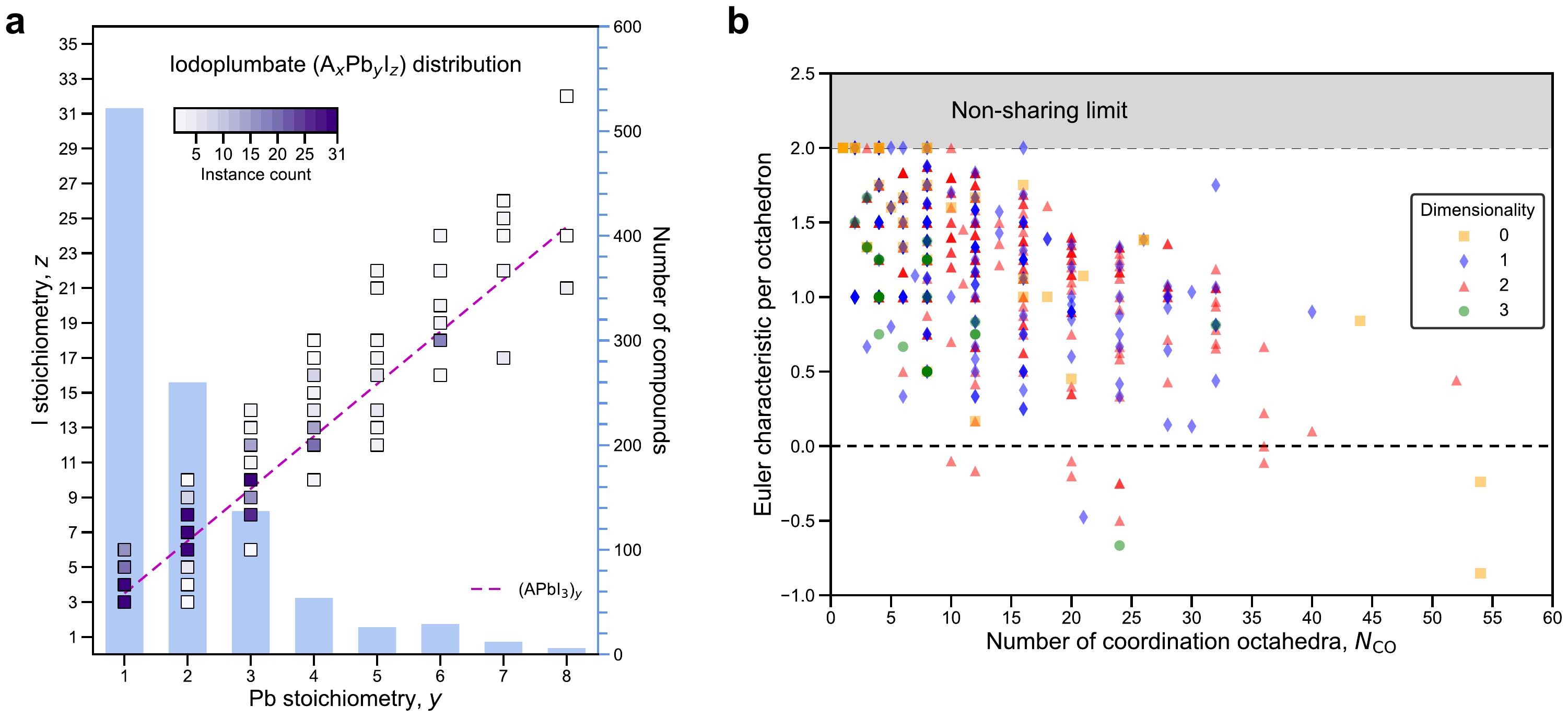}
    \caption{\textbf{Landscape of iodoplumbate octahedral networks.} \textbf{a}, Existing stoichiometries of hybrid iodoplumbates (A$_x$Pb$_y$I$_z$) mined from the Cambridge Structural Database. \textbf{b}, Relationship between the number of coordination octahedra ($N_{\mathrm{CO}}$) and the Euler characteristic per octahedron ($\chi^{\mathrm{ON}} / N_{\mathrm{CO}}$). Each point represents an example of hybrid iodoplumbates, colored according to the framework dimensionality \cite{Guinier1984} of the material.}
    \label{fig:iodoplumbates}
  \end{center}
  \vspace{-1.5em}
\end{figure*}

Previous attempts at linking stoichiometry to dimensionality \cite{Mercier2009,Wu2009} showed limited success in small case studies \cite{Qian2017,Deng2018} and became less effective with the increasing complexity of the materials design space. In our view, the limitation is due to the restricted feature space and problem formulation (i.e., predicting dimensionality using stoichiometry). A taxonomy should also distinguish the topological aspect of the structural components \cite{Blatov2004}, which is equivalent to polymorph (or polytype) identification. Our procedure builds on these requirements for motif assignment in hybrid metalates, which aims to distinguish the inorganic polytypes, regardless of the organic constituent. To this end, we analyzed $\sim$ 970 compounds from the Cambridge Structural Database that are hybrid iodoplumbates (see Methods), A$_x$Pb$_y$I$_z$, where A is an organic cation, and the [Pb$_y$I$_z$]$^{n-}$ polynuclear anion induces polytypism. The hybrid iodoplumbates are the most studied members of the hybrid metalate materials family that also includes hybrid perovskites \cite{Mercier2009,Wu2009,Jin2021}.
\begin{figure*}[htbp!]
  \begin{center}
    \includegraphics[width=0.98\textwidth]{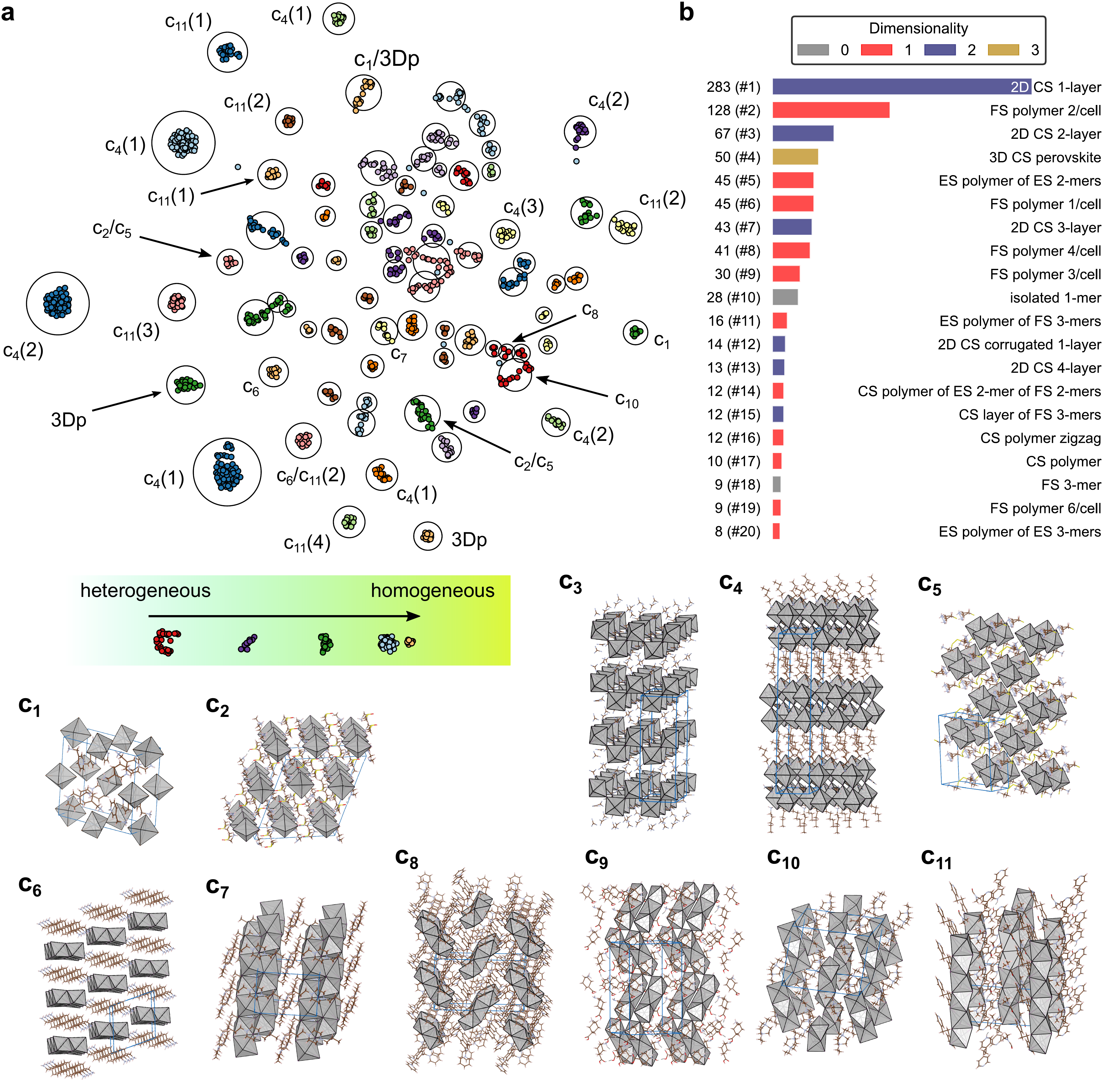}
    \caption{\textbf{Computational taxonomy of octahedral networks in A$_x$Pb$_y$I$_z$.} \textbf{a}, Embedding space visualization of Pb$_y$I$_z$ octahedral networks. Some of the major clusters are labeled with shorthand notations in connection to \textbf{c}. The number in parentheses indicates the number of building units per unit cell. The black circles and the colors distinguish the identified clusters from unsupervised machine learning. \textbf{b}, The top twenty polytypes in the occurrence ranking derived from the clustering in \textbf{a} and human label refinement. The colors in the bar chart represent the grouping by dimensionality. The representative structures are shown in \textbf{c$_1$}-\textbf{c$_{11}$}, from monomeric to polymeric octahedral networks with diverse motifs. The names of the motifs are \textbf{c$_1$}, non-sharing 1-mer (0D perovskite), \textbf{c$_2$}, straight CS 2-layer; \textbf{c$_3$}, corrugated CS 1-layer; \textbf{c$_4$}, straight CS chain; \textbf{c$_5$}, corrugated CS chain; \textbf{c$_6$}, ES chain of ES 2-mers; \textbf{c$_7$}, CS chain of ES 2-mer of FS 2-mers; \textbf{c$_8$}, isolated FS 3-mers; \textbf{c$_9$}, CS layer of FS 3-mers; \textbf{c$_{10}$}, ES chain of FS 3-mers; \textbf{c$_{11}$}, FS chain. In terms of framework dimensionality, polytype \textbf{c$_1$} and \textbf{c$_8$} are 0D, polytype \textbf{c$_2$}, \textbf{c$_5$}, \textbf{c$_6$}, \textbf{c$_7$}, \textbf{c$_{10}$}, \textbf{c$_{11}$} are 1D, polytype \textbf{c$_3$}, \textbf{c$_4$}, \textbf{c$_9$} are 2D. The solid blue box outlines the unit cell of the respective structure.}
    \label{fig:prototypes}
  \end{center}
\end{figure*}

\textbf{Structure mining of A$_x$Pb$_y$I$_z$}. We first explored the stoichiometric and topological diversity of A$_x$Pb$_y$I$_z$ compounds using our computational workflow. In Fig. \ref{fig:iodoplumbates}a, the stoichiometric ratio between lead (Pb) and iodine (I) in the empirical formula is on the order of 3:1, yielding an average formula of (APbI$_3$)$_y$. As the number of Pb atoms in the formula unit increases, so does the scatter along the magenta dashed line representing $z\sim3y$, indicating stoichiometries that more frequently deviate from this 3:1 ratio due to the growing complexity of network connectivity. When representing the polynuclear anions ([Pb$_y$I$_z$]$^{n-}$, with $n$ being the charge it bears) by 3D meshes, we can calculate the relationship between the number of COs and the Euler characteristic \cite{Klain1997,Naskrecki2022} of the mesh graph representing the ONs (see SI section 1). The results in Fig. \ref{fig:iodoplumbates}b lie within a region bounded by the non-sharing limit, where isolated octahedra (i.e., 0D hybrid perovskites \cite{Sun2021_0D}) exist. The Euler characteristic is an invariant metric for topological objects. For convex shapes such as single convex octahedra, it attains a value of 2, while it becomes less than 2 for nonconvex shapes \cite{Klain1997,Naskrecki2022}. Fig. \ref{fig:iodoplumbates}b shows that as the number of COs increases, the Euler characteristic generally decreases in value, indicating a departure from convex shapes, as a result of atom-sharing between neighboring COs. This observation supports the use of topological features to distinguish between inorganic polytypes.

\begin{figure*}[htbp!]
  \centering
    \includegraphics[width=0.98\textwidth]{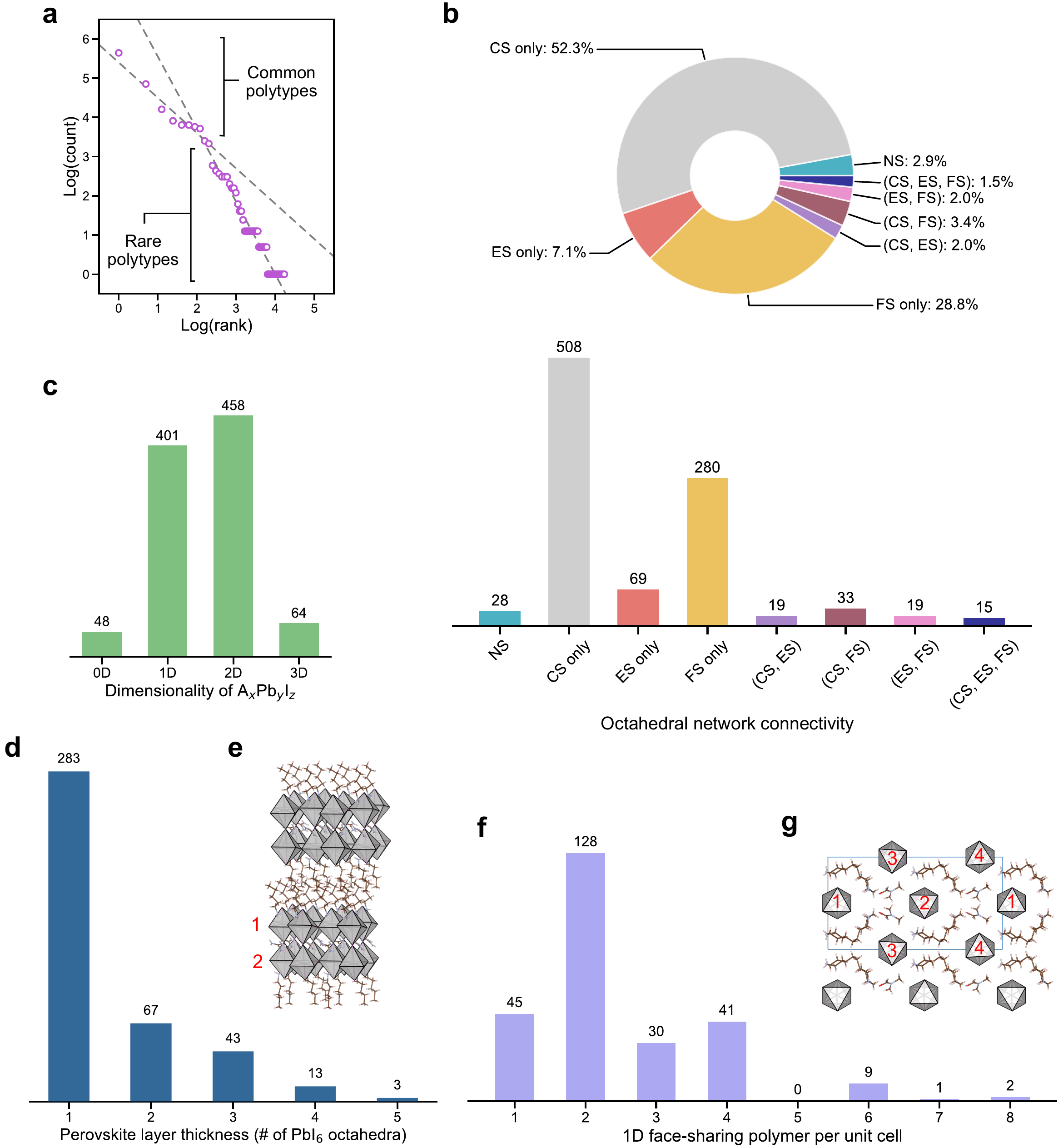}
    \caption{\textbf{Statistical summary of octahedral networks in A$_x$Pb$_y$I$_z$.} \textbf{a}, The ranking distribution of the inorganic polytypes in the A$_x$Pb$_y$I$_z$ dataset contains distinct regimes for common and rare polytypes in a log-log plot. \textbf{b}, Distribution of the structures differentiated by their PbI$_6$ octahedral connectivity types. Here, CS (corner-sharing) only, ES (edge-sharing) only, and FS (face-sharing) only include the structures with the respective unary connectivity type. NS indicates non-sharing. The structures containing mixed connectivity are divided into binary and ternary connectivity types and labeled with the constituent types in parentheses. \textbf{c}, Framework dimensionality of the structures. \textbf{f}, Distribution of the layer thickness in 2D layered perovskites (polytype name CS $k$-layer, $k$ = 1, 2, ...), along with an example structure in \textbf{e} showing a CS 2-layer polytype. \textbf{f}, Distribution of 1D FS chains per unit cell, along with an example structure in \textbf{g} showing four symmetry-independent FS chains per unit cell, which is indicated by the solid blue box.}
    \label{fig:pbi_stats}
    \vspace{-0.5em}
\end{figure*}

\textbf{Clustering with CNE representation}. The CNE is a vector representation of the ON constructed with the stoichiometric and topological information of the Pb$_y$I$_z$ motif (see SI section 3). We perform dimensionality reduction and clustering interactively to identify the distribution of polytypes, as shown in Fig. \ref{fig:prototypes}a. Manifold learning is used here for dimensionality reduction because it provides an interpretable (due to well-separated clusters) visualization \cite{bibal2019safeml} of underlying structures of the dataset to facilitate subsequent clustering. We validate the clusters derived from the algorithms using both external and internal metrics, along with polytype labels generated by visualization and human refinement (see Tables S3-S6). The clusters within the A$_x$Pb$_y$I$_z$ dataset \cite{Hennig2015} uncovered here correspond to structures with a shared ON motif. Details of the procedure are provided in SI section 3. We rank the inorganic polytypes according to their occurrence in the dataset, revealing a double Pareto behavior (i.e., power-law relation involving two regimes) \citep{mitzenmacher2003}, as in Fig. \ref{fig:pbi_stats}a, and the top-twenty entries (out of $\sim$ 70) are listed in Fig. \ref{fig:prototypes}b. The existence of two power-law coefficients indicates distinct regimes. Here, they are most likely a result of the selective research activities associated with these materials that distinguish the common from the rare polytypes. The common polytypes are chemically fine-tuned to improve their key properties. Whereas the rare polytypes appear in unique studies or are characterized as intermediate products in chemical fine-tuning \cite{Daub2021,nellikkal_spatiotemporal_2022,gilley2025}. Some of the prominent polytypes are visualized by their examples in Fig. \ref{fig:prototypes}c.

We find that a number of clusters already show high purity in Fig. \ref{fig:prototypes}a, where homogeneous clusters tend to have a more globular shape than heterogeneous ones. The shorthand notations in \ref{fig:prototypes}a correspond to the indices in Fig. \ref{fig:prototypes}c and its potential variants. For example, $\text{c}_4(1)$ indicates CS 1-layer polytype, or single-layer 2D hybrid perovskite, $\text{c}_{11}(2)$ indicates having the FS chain polytype with two independent chains per unit cell, 3Dp represents 3D CS hybrid perovskites. Clusters with two major polytypes are indicated with a slash, such as $\text{c}_2/\text{c}_5$. A complete description of the naming rules for the polytypes is provided in Methods. More examples for the identified inorganic polytypes are provided in Table S5.

\begin{figure}[htb!]
  \centering
    \includegraphics[width=0.46\textwidth]{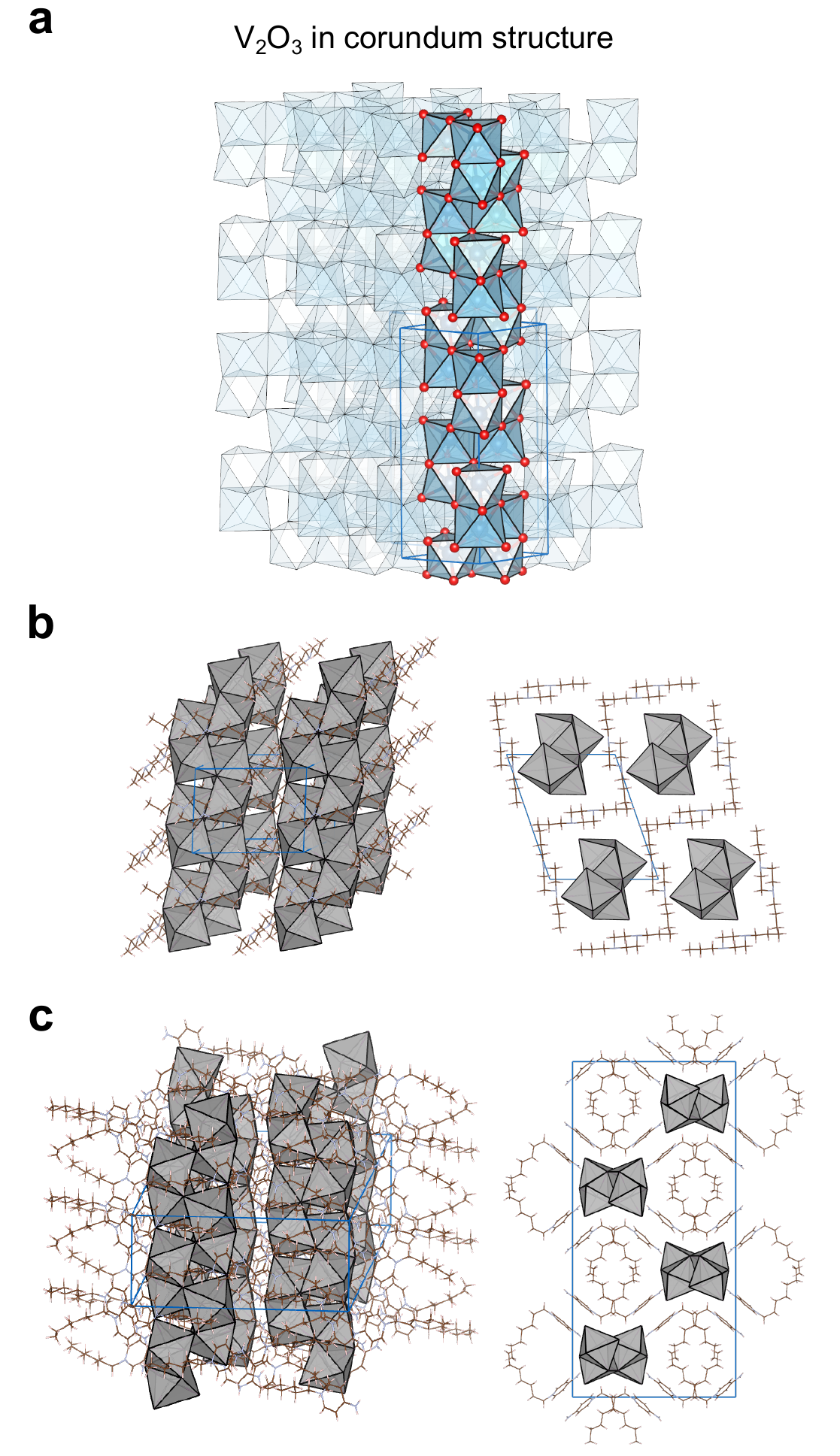}
    \caption{\textbf{Corundum-like columns in mixed-connectivity} A$_x$Pb$_y$I$_z$. \textbf{a}, Illustration of the mixed connectivity columns (including corner-, edge-, and face-sharing simultaneously) in the corundum structure using V$_2$O$_3$ as an example. The cyan octahedra has the composition VO$_6$ and the oxygen atoms are in red. Similar columns with mixed connectivity exist in hybrid iodoplumbates, as in the examples \textbf{b}, (C$_{10}$H$_{24}$N$_2$O)$_2$Pb$_4$I$_{10}$ \cite{hou2012} and \textbf{c}, (C$_{12}$H$_{21}$N$_2$)$_2$Pb$_2$I$_6$ \cite{duan2016}, both shown in their respective side (left) and top (right) views. The blue box indicates the unit cell in all views.}
    \label{fig:corundum}
    \vspace{-0.5em}
\end{figure}

The major clusters in Fig. \ref{fig:prototypes}a represent common network motifs: CS $k$-layers (of different layer thickness $k$), FS chains (with different numbers of polymeric chains per unit cell), 0D and 3D hybrid perovskite structures. Some of the lesser-known polytypes are also aggregated into high-purity clusters, including those with mixed connectivity such as ES chain of FS 3-mers (Fig. \ref{fig:prototypes}c$_{10}$), CS layer of FS 3-mers (Fig. \ref{fig:prototypes}c$_9$), and CS chain of ES 2-mer of FS 2-mers (Fig. \ref{fig:prototypes}c$_7$), which has a 1D arrangement of COs similar to the columnar substructure of the V$_2$O$_3$ corundum (see Fig. \ref{fig:corundum}) \cite{hou2012,duan2016}. Although the mixed connectivity in this polytype due to the [Pb$_2$I$_6$]$^{2-}$ anion is in seeming violation of Pauling's third rule, it's surprisingly more common than many other inorganic polytypes found in the A$_x$Pb$_y$I$_z$ dataset. A caveat to mention is that the clusters in Fig. \ref{fig:prototypes}c with the same label generally differ in the packing of the organic cation and its charge ($n/x$ for [A$_x]^{n+}$), which is not considered explicitly in polytype identification but emerges due to the correlation between the characteristics of the organic cation and the inorganic framework it supports. Chemically speaking, the organic packing is a part of the structural motifs of these materials, but is not the network motif of the PbI$_6$ COs.

Moreover, we compare the effectiveness of the CNE with an established atom-centered representation, the smooth overlap of atomic positions (SOAP) \cite{Bartok2013}, a kernel-based symmetry-invariant representation that prioritizes distance information, which has been used to visualize materials design spaces \cite{Cheng2020}. The results from our controlled numerical experiments (see Fig. \DRtuning\, and Table S4) show that, because of the scale dependence of the kernel, clusters from the SOAP representation \cite{Bartok2013} are generally less pure than those obtained from the CNE scheme, and exhibit more elongated shapes, which are less beneficial for the subsequent polytype identification.

\textbf{Pauling-like rules for iodoplumbates}. The statistical summary of identified polytypes shows that in hybrid iodoplumbates, a significant proportion of existing materials contain only CS or FS octahedra, while some also contain CS octahedra co-existing with FS or ES octahedra in the same structure (Fig. \ref{fig:pbi_stats}b). The nearest-neighbor connectivity pattern of COs has the trend,
\begin{equation}
% \begin{split}
    \text{corner-sharing PbI}_6 > \text{face-sharing PbI}_6 \gg \text{edge-sharing PbI}_6.
% \end{split}
\label{eq:nonpauling}
\end{equation}
In terms of probability of connectivity ($P$), we can equivalently write $P_{\mathrm{PbI}_6}^{\mathrm{CS}} > P_{\mathrm{PbI}_6}^{\mathrm{FS}} \gg P_{\mathrm{PbI}_6}^{\mathrm{ES}}$. This observation violates Pauling's third rule \cite{Pauling1929} for the structure prediction of ionic compounds, which orders corner-sharing (CS) as the most likely connectivity type between neighboring coordination environments, followed by edge-sharing (ES), and then face-sharing (FS). The violation leads to a modified preference in connectivity, which is observed in layered oxides \cite{Delmas1980}, oxides under non-ambient conditions \cite{Bykova2018}, and oxides with large anion size and high coordination number \cite{George2020}. However, for halides, no systematic study is yet present.

For structural prediction, Pauling's third rule still provides a good estimate for the \textit{dominant} connectivity type between coordination environments,\cite{George2020,Pathak2025} which is CS in our case for a homogeneously connected coordination network (i.e., containing only a single connectivity type). For those with two distinct connectivity types, our summary in Fig. \ref{fig:pbi_stats}b) shows that the polytype with coexisting CS and FS PbI$_6$ octahedra, the two main connectivity types in Eq. \eqref{eq:nonpauling}, is the most common, followed by the other two possibilities,
\begin{equation}
% \begin{split}
    \text{corner- \& face-sharing PbI}_6
    \gg \text{corner- \& edge-sharing PbI}_6
    \sim \text{face- \& edge-sharing PbI}_6.
% \end{split}
\label{eq:pauling_2nd}
\end{equation}
This may also be written more compactly in the probability of connectivity as $P_{\mathrm{PbI}_6}^{\mathrm{CS, FS}} \gg P_{\mathrm{PbI}_6}^{\mathrm{CS, ES}} \approx P_{\mathrm{PbI}_6}^{\mathrm{FS, ES}}$, and $P_{\mathrm{PbI}_6}^{\mathrm{CS, FS}} \propto P_{\mathrm{PbI}_6}^{\mathrm{CS}}P_{\mathrm{PbI}_6}^{\mathrm{FS}}$, similarly for the other two probabilities for mixed connectivity. Eq. \eqref{eq:pauling_2nd} may be regarded as the second-order extension of Pauling's third rule \cite{Pauling1929} for the specific type of coordination environment encountered here.

\textbf{Structure occurrence and design principles}. The polytype occurrence ranking in Fig. \ref{fig:prototypes}b reveals more information supplementary to the connectivity trends discussed in Eqs. \eqref{eq:nonpauling}-\eqref{eq:pauling_2nd}. By cross-examination, we find a scarcity of structures with 2D layers entirely composed of edge-sharing PbI$_6$ octahedra (only 3 out of $\sim$ 970). In contrast, the 2D edge-sharing motif is the majority polytype of PbI$_2$ \cite{Beckmann2010}, a parent compound of A$_x$Pb$_y$I$_z$, as well as the structure of the ``analogous perovskite'' compound BiI$_3$ (in the sense that ReO$_3$ is structurally analogous to ABO$_3$) \cite{Evans2020} that has been reported \cite{Brandt2015}. This is likely due to the balance of charges since infinite edge-sharing PbI$_6$ octahedral layers are charge neutral and would require vacancies to bear charges \cite{Saparov2016} and incorporate $A$-site cations. In fact, the more common types of ONs in A$_x$Pb$_y$I$_z$ with edge-sharing connectivity are predominantly 1D, featuring a short secondary unit, such as an edge-sharing 2-mer or 3-mer, stacked up through corner-sharing or edge-sharing (e.g., Fig. \ref{fig:prototypes}c$_6$). Impressively, the 1D multimeric polytypes with a single connectivity type (i.e., only containing CS, ES, or FS octahedra), traditionally found in materials with distinct oxidation states \cite{Rouxel1986}, are all present within this single family of materials via pairing with different organic cations, indicating the tremendous framework variability of hybrid materials. These observations suggest that polyatomic organic cations can stabilize more diverse ONs than those typically found in the (parent) inorganic materials, where only monatomic cations exist.

In addition to rationalizing the observed trends, examination across polytypes with similar building units helps identify design principles related to (i) dimensionality tuning by changing connectivity, and (ii) building unit augmentation. For principle (i), we find that the dimensionality may be tuned by the octahedral connectivity within the inorganic framework. This enables the use of ES or FS multimers to construct, in principle, ONs of any dimension (0D-3D), greatly expanding the perovskite motif with single CS octahedra. Although dimensional tuning has been discussed in the context of purely inorganic materials \cite{Tulsky2001}, for hybrid metalates such as iodoplumbates, the vast space of admissible organic cations \cite{karimitari2024} suggests the possibility of fine-tuning the material properties by changing the inorganic framework. For principle (ii), we find that accommodation of the ON to different organic cations may also proceed by augmenting the secondary unit, such as from 1-mers or 2-mers to 6-mers, which are all found in the A$_x$Pb$_y$I$_z$ dataset. Besides (unidirectional) stacking, which defines polytypism \cite{Guinier1984}, circular arrangements stand out as another distinguishing spatial construction \cite{krautscheid1999}. The two principles outlined here are further discussed and illustrated with examples in Figs. \fcontinuum-\fmodulardesign.

\section{Discussions}
From a computational perspective, our approach marries chemical intuition with geometry processing and the landscape of machine learning methods using object-centric representations \cite{Grohe2020}. The objects here refer to the atom-centered COs present in the source data, rather than the atoms reported from simulation or experiment. Because of this object focus, the outcome is highly interpretable. The present work motivates further developments in efficient feature extraction methods. For example, existing approaches often require a symmetry-checking step \cite{Ganose2019,Waroquiers2020} that becomes computationally expensive for large datasets of complex structures. This necessitates simpler procedures followed by data cleaning using chemical intuition, which produces noisier data (e.g., with missed or inaccurately identified coordination environments). As such, the scalability of our approach is heavily dependent upon the efficiency and accuracy of coordination environment detection techniques.

From a materials chemistry perspective, our oxide and iodoplumbate results deserve further contemplation with regard to their published energetics and related material properties. Dimensional confinement effects \cite{Hoye2022}, which stem from changes in octahedral connectivity, significantly modify the electronic bandgap and optoelectronic properties in hybrid iodoplumbates and related ternary halides \cite{Kamminga2017,Stoumpos2017,Mauck2019,liu_highperf_2023}. In halide compositions, for instance, purely ES connectivity increases the bandgap by $\sim$ 0.5 eV over purely CS connectivity, and FS increases the bandgap by another $\sim$ 0.5 eV \cite{Kamminga2017}. This is comparable to the bandgap broadening observed from compositional tuning by switching from iodide to bromide anions ($\sim$ 0.5 eV) and bromide to chloride anions ($\sim$ 0.6 eV) \cite{Marchenko2020}. Both effects are related to the reduced overlap of the atomic orbitals due to the lower dimensionality or the smaller anions. Moreover, the electronic dimensionality affects the photoluminescence properties of iodoplumbates: ES and FS motifs can lead to charge localization, which can enhance the photoluminescence quantum yield up to 100\% \cite{Jin2021}. On the contrary, CS and mixed-connectivity motifs can be beneficial for charge transport and modified emissions (e.g., broad emission \cite{Mao2017}). For multinary hybrid metalates, the dimensionality and connectivity of their octahedral networks can also be altered by their synthesis conditions \cite{Daub2021}.

\section{Conclusion and outlook}
Our results demonstrate the advantage of task-dependent representations for machine learning-assisted exploration of the materials design space \cite{Onat2020} and the need to explicitly incorporate connectivity information in existing distance-based representations \cite{Bartok2013,Banjade2021}. The diversity of ON motifs cataloged in the present work motivates further investigations into the relationship between connectivity patterns and application-specific materials properties \cite{Dyer2013,Park2021} computed with density functional theory. The CNE representation may be merged with those for organic molecules to build interpretable predictors and generative models of hybrid functional material properties. The modified Pauling's third rule in the materials examined here and by others \cite{Delmas1980,Bykova2018,George2020,Pathak2025} motivates its reformulation using a fully probabilistic approach that reflects chemical specificity and serves growing data. We envision a computational study whereby the functionalities could be tuned for a fixed ON, with pairing moieties that are optimized in an approach that resembles the docking mechanism in protein chemistry. Insights from such a study would facilitate targeted screenings for specific network connectivity patterns \cite{Park2021}, and likely reduce the resource cost and attrition rate from random search-based screening methods \cite{Oganov2019}. It will grant us a broader accessible space using computational structure-prediction methods to systematically investigate new phases as well as material properties associated with connectivity modifications.

In addition to the materials studies in this work, our approach can also be adapted to extract insights from other analogous, yet less investigated materials families featuring ONs. They include inorganic and hybrid metalates with nonmetal or metalloid ligands, high-entropy materials \cite{Musico2020}, coordination polymers, and molecular perovskites \cite{Bostrom2020}, which possess even more complex design spaces and highly tunable structure-property landscapes. Our approach reflects the ongoing shift in materials design from an atom-centric perspective to one focused on coordination environment, which can capture the hierarchical organization of complex materials that gives rise to long-range order. This modular perspective supports the rational design of multinary and molecular perovskites, where collective polyhedral distortions dictate emerging optoelectronic functionality \cite{Bostrom2020,gilley2025}.

\section*{Methods}
\setstretch{1.}
\small
\subsection{Materials datasets and their curation}
The oxide perovskite (ABO$_{3}$) polymorph dataset was obtained from high-throughput computational screening of compositionally valid compounds. We provide here a summary of the computational method with further details reported elsewhere \cite{Bare2022}: The structures of the oxide perovskites were generated using the software SPuDS \cite{Lufaso2001}, which accounts for electrostatic structural distortions using bond-valence theory. The SPuDS-generated structures were further optimized using density functional theory within the Vienna Ab-initio Simulation Program (VASP 5.4.1) and under periodic boundary conditions utilizing projector augmented wave (PAW) pseudopotentials and the Perdew-Burke-Ernzerhof (PBE) generalized gradient approximation (GGA) exchange-correlation functional \cite{Perdew1996}. Electronic self-consistent field energies were converged to within $10^{-6}$ eV and forces were converged to within 0.01 eV/$\text{\AA}$. The dataset is represented in the \texttt{pymatgen}-style json format \cite{Ong2013}.

The structural data for hybrid iodoplumbates (A$_x$Pb$_y$I$_z$) were obtained from the Cambridge Structural Database (CSD) \cite{Groom2016}, accumulated until early 2022. We directly queried the CSD via a chemical search of lead and iodine compounds, which yielded over 1020 entries. We next removed entries that were not hybrid iodoplumbates, including (i) structures that did not contain organic components (removed by checking the chemical species beyond Pb and I in the compound) and (ii) structures with lead coordination environments other than octahedral (removed during the initial coordination environment detection). We also kept only the compounds with a Pb stoichiometry of up to 6 in the chemical formula. The final number of stoichiometrically-ordered data entries was 971, which includes fewer than 6 entries with duplicated IDs in the CSD. These duplicates were kept because of their different measurement conditions.

The majority of data processing was carried out using \texttt{crystmorph}, which implements the workflow described in SI section 1 for the analysis of ONs. We adopted a chemical environment parser \cite{Waroquiers2020} to analyze the structures by iteratively identifying the space groups of possible coordination environments centered locally on every atom. For the structures used in this work, the analysis was performed only for metal-centered coordination environments to speed up the process. The coordination environment is partitioned into COs, which are the base objects in \texttt{crystmorph}. The algebraic operations between COs, as described in SI section 1, are implemented at either the atomic level (mesh graph) or the CO-level (inter-unit graph).

\subsection*{Naming rules for inorganic polytypes}
There exist a few naming systems for modular (or recombination) materials, largely based on the fragments \cite{Guinier1984,Lima-de-Faria1990}. For 3D frameworks, the notations from Ramsdell \cite{Li_evolutionary_2021} and Jagodzinski \cite{Kunz2024} for polytypes have been used, yet they are limited to layer-stacking structures with a single chemical composition. These notations cannot concretely capture non-stacking structures (e.g., 0D structures) \cite{Sun2021_0D} comprised of atom-centered COs with shared chemical origins as the parent compound. Moreover, hybrid materials contain distinct chemical components (e.g., inorganic and organic) and exhibit increased structural diversity from sublattice composition and mixing \cite{Bartel2020,wiggins2025}. Their description goes beyond the naming systems designed for purely inorganic (binary) compounds \cite{Wells1977,Muller1981,Burdett1982}. The polytypism is only interpreted at the level of the inorganic sublattice, \cite{gilley2025} hence inorganic polytypes. For hybrid iodoplumbates (and similarly for other halometalates), we can treat the [Pb$_y$I$_z]^{n-}$ part as a polyatomic ion derived from a polytypic binary compound, which forms the inorganic framework. The naming procedure is divided into the following rules.
\begin{enumerate}[wide, labelindent=10pt]
\item[1.] \textbf{Degree of connectivity.} The \textit{degree of connectivity} is defined as the number of shared vertices between two octahedra, which could be 0, 1, 2, or 3, for non-sharing (NS), corner-sharing (CS), edge-sharing (ES), or face-sharing (FS), respectively. The naming starts from the octahedral units with the highest degree of connectivity and proceeds in decreasing order.

\item[2.] \textbf{Building units.} In atomic ONs, the \textit{primary (building) units} of the structure are the COs with atomic ligands, the \textit{secondary (building) units} are constructed from primary units through \textit{gluing} operations \cite{Gallier2008} such as atom-sharing (i.e., corner, edge, face-sharing), which form \textit{multimers}. Similarly, a \textit{tertiary (building) unit} can be constructed from gluing secondary units through corner-, edge-, or face-sharing, or a mixture of them.

\item[3.] \textbf{Multiplicity of building units.} This step starts from identifying the smallest and most connected secondary unit (a multimer) within the ON. In graph-theoretical terms, this amounts to the maximally connected subgraph of the inter-unit graph representing the ON. Then, determine the \textit{multiplicity} of the building units. For a multimeric structure, the multiplicity is the number of COs involved in the secondary unit. A \textit{multimer} (or $k$-mer) has a secondary unit with a multiplicity of $k$ ($k$ is a positive integer). Naturally, a dimer, trimer, tetramer, and pentamer may be referred to as 2-mer, 3-mer, 4-mer, and 5-mer, respectively. Two limiting cases of the multiplicity exist: The lower limit ($k=1$) is a \textit{monomer}, such as the isolated 1-mers (0D structures)\cite{Sun2021_0D} in Fig. \ref{fig:prototypes}. The upper limit ($k=\infty$) is a \textit{polymer} or \textit{polymeric chain} of octahedra, which is equivalent to an $\infty$-mer (infinite-mer).

\item[4.] \textbf{Elementary and complex multimers.} A multimer can be corner-, edge-, or face-sharing, or have a mixture of them. They are called CS multimer, ES multimer, FS multimer, or mixed-connectivity multimer, respectively. An \textit{elementary multimer} has only one type of connectivity, while a \textit{complex multimer} has at least two types of connectivity.

\item[5.] \textbf{Decorated multimers.} A multimer can be \textit{linear} (1D) or, equivalently, \textit{flat} (2D), zigzag (1D) or, equivalently, \textit{corrugated} (2D), \textit{curved}, or \textit{circular}, which are decorative terms to illustrate the fine features at larger spatial scales, typically requiring at least a 3-mer unit to represent. A straight multimer is the default type, so the adjective is dropped in its name, but for other types, the adjective is necessary. They are the corrugated $k$-mer, curved $k$-mer, or circular $k$-mer, respectively. In a corrugated multimer, the octahedral units are connected consecutively on alternating sides of the previous one \cite{Mao2017}. In a curved multimer, the octahedral units are getting progressively closer to the initial one, but are not forming a closed circular pattern \cite{mishra2008}. In a circular multimer, the terminal octahedral unit forms a closure with the initial one through atom-sharing \cite{krautscheid1999,yue2019}.

\item[6.] \textbf{Framework dimensionality.} An octahedral framework can be polymeric (1D \textit{chain} or polymer), layered (2D \textit{layer}), or scaffolded (3D \textit{framework}). This level of specification naturally yields the dimensionality of the framework \cite{Lima-de-Faria1990,Hoffmann2013}.

\item[7.] \textbf{Naming sequence.} The name of the inorganic framework is constructed sequentially from the primary unit (U$_1$) to the secondary unit (U$_2$) and above. The name is written as ``U$_m$ of U$_{m-1}$ of ... of U$_2$ of U$_1$'', with increasing hierarchy from U$_1$ to U$_m$.
\end{enumerate}

The set of naming rules we propose here is meant for generating searchable polytype names and calculating summary statistics for a structure collection. They are based on the ON motifs and are constructed intuitively to account for the multiscale nature of the connectivity pattern present in the material structures. The polytype names produced from these rules contain explicit information of the inorganic framework, so are also usable as labels for machine learning tasks, such as predicting the connectivity types, the framework dimensionality, the multiplicity of building units, conditioned on the pairing moieties. The use of these naming rules is illustrated from the ground up with two real examples in Fig. S7. More examples of the named polytypes are given in Table S5.

We note here that these naming rules will not be able to cover all possible CO combinations or ON architectures, as their variability can grow exponentially, nor can any of the conventions that have been proposed so far in the crystallography community \cite{Guinier1984,Lima-de-Faria1990,Muller2007,Kunz2024}. However, our naming rules can handle the vast majority of cases for halometalates (and likely for related metalates) involving ONs and contain a level of generality within their designed scope.

% \begin{suppinfo}
\section*{Data availability statement}
\setstretch{1.2}
\normalsize
The single oxide perovskite polymorph dataset (structures) is available on GitHub (\href{https://github.com/rymo1354/gii\_minimization/tree/main/structures\_and\_energies}{link}). The hybrid iodoplumbate dataset (features, results, and IDs of Cambridge Structural Database) is provided in the supporting information.

% Supporting Information: 
\section*{Supporting information}
\begin{itemize}
  \item Extended details of the two structure analysis case studies (oxides and iodoplumbates), theoretical calculations of oxidation states, description of the computational workflow, and named examples for the inorganic polytypes illustrating their trends (PDF).
  \item Table S5: A selection of inorganic polytypes of A$_x$Pb$_y$I$_z$ ranked by occurrence (PDF).
  \item Table S6: Results from unsupervised classification along with human-refined labels (XLSX).
\end{itemize}

% \end{suppinfo}


\section*{Note}
The authors declare no competing financial interest. The code for the analysis of octahedral networks is accessible on GitHub (\href{https://github.com/RealPolitiX/crystmorph}{https://github.com/RealPolitiX/crystmorph}).

% \begin{acknowledgement}

\section*{Acknowledgements}
We thank G. Cs\'{a}nyi (University of Cambridge) and H.-C. zur Loye (University of South Carolina) for helpful discussions. The work was supported by the Air Force Research Laboratory through the research grant (FA8655-21-1-7010). C.B.M. and R.J.M. were supported by the U.S. Department of Energy (DOE), Office of Energy Efficiency and Renewable Energy (EERE), Hydrogen and Fuel Cell Technologies Office (HFTO), and specifically the HydroGEN Advanced Water Splitting Materials Consortium, established as part of the Energy Materials Network under this same office (award DE-EE0008088). C.B.M. and R.J.M. also acknowledge support from the National Science Foundation (awards NSF CHEM-1800592 and CBET-2016225). I.H. acknowledges support from the Israel Science Foundation (Grant No. 2078/23). C.S. acknowledges start-up support from the University of South Carolina.

% \end{acknowledgement}
\bibliography{achemso-refs}

% \newpage
% \begin{figure*}
%     \centering
%     \includegraphics[width=\linewidth]{Figures/ChemMater_TOC.png}
%     \begin{flushleft}
%         Highlights of the two case studies in this work on the computational analysis of materials distinguished by their octahedral networks. Left: Axis-resolved tilting trends from A-site substitution in lanthanide oxide perovskite series (LnBO$_3$). Right: Pauling-like rules for the prevalence of connectivity patterns among the inorganic polytypes of hybrid iodoplumbates (A$_x$Pb$_y$I$_z$).
%     \end{flushleft}
%     \label{fig:toc}
% \end{figure*}

\end{document}